\documentclass[12pt]{article}
\usepackage{upgreek}
\usepackage{mathrsfs}
\usepackage{amsmath,amssymb,color}
\usepackage{amsfonts}
\usepackage{url}
\usepackage{xcolor}
\usepackage[colorlinks,linkcolor=black,anchorcolor=blue,citecolor=blue,urlcolor=blue]{hyperref}
\usepackage{graphicx}
\usepackage{wrapfig}
\usepackage{cases}
\usepackage{appendix}
\usepackage[T1]{fontenc}
\usepackage{authblk}
\usepackage{txfonts}
\usepackage[numbers,sort&compress]{natbib}
\makeatletter
\@addtoreset{equation}{section}
\makeatother

\newcommand{\er}[1]{\eqref{#1}}

\newcommand{\ci}[1]{}

\newcommand{\ke}{\rangle}
\newcommand{\br}{\langle}
\newcommand{\lb}{\left(}
\newcommand{\rb}{\right)}

\newcommand{\lsb}{\left[}
\newcommand{\rsb}{\right]}

\newcommand{\nn}{\nonumber \\}
\newcommand{\p}{\partial}

\newcommand{\ba}{\begin{eqnarray}}
\newcommand{\ea}{\end{eqnarray}}
\newcommand{\be}{\begin{equation}}
\newcommand{\ee}{\end{equation}}
\newcommand{\bal}{\begin{align}}
\newcommand{\eal}{\end{align}}
\newcommand{\bay}[1]{\left(\begin{array}{#1}}
\newcommand{\eay}{\end{array}\right)}
\newcommand{\eg}{\textrm{e.g.} }
\newcommand{\ie}{\textrm{i.e.}, }
\newcommand{\iv}[1]{{#1}^{-1}}
\newcommand{\st}[1]{|#1\ke}
\newcommand{\at}[1]{{\Big|}_{#1}}
\newcommand{\zt}[1]{\textrm{#1}}

\def\rmd{{\rm d}}
\def\xa{{\alpha}}

\def\xd{{\delta}}
\def\xD{{\Delta}}
\def\xe{{\epsilon}}

\def\xg{{\gamma}}

\def\xs{{\sigma}}
\def\xS{{\Sigma}}

\def \Tr {{\rm Tr}}

\def\CF{{\cal F}}

\def\CH{{\cal H}}

\def\CK{{\cal K}}

\def\CO{{\cal O}}

\def\CW{{\cal W}}

\def\BH{\mathbb{H}}

\def\h{{\frac 1 2}}

\title{A note on the kinematic space associated with a subregion}
\author[a,b,c]{Xing Huang \thanks{xingavatar@gmail.com}}
\affil[a]{Institute of Modern Physics, Northwest University, Xi'an 710069, China}
\affil[b]{Shaanxi Key Laboratory for Theoretical Physics Frontiers, Xi'an 710069, China}
\affil[c]{NSFC-SPTP Peng Huanwu Center for Fundamental Theory, Xi'an 710127, China}

\date{}
\begin{document}

\maketitle
\begin{abstract}
We discuss the issue in constructing the kinematic space of the geodesics lying partially inside the entanglement wedge associated with a single interval. We then resolve the problem by working with the equivalent kinematic space of the reflected geodesics. We also show that the length of a reflected geodesic corresponds to the (generalized) reflected entropy, which can be computed using entirely the information obtained from the reduced density matrix, satisfying the requirement of the subregion-subregion duality.

\end{abstract}

\newpage
\noindent\rule[0.1\baselineskip]{\textwidth}{0.8pt}
\tableofcontents
\noindent\rule[0.1\baselineskip]{\textwidth}{0.8pt}
\linespread{1.5}\selectfont

\section{Introduction}
The intermarriage between quantum information and gravity has proved to be a prolific approach, which may get around the intricacies of quantum gravity and
reveal some of its universal features. Information theoretic quantities could emerge as seemingly
very different geometric objects. For example in our favorite testing ground
of AdS/CFT \cite{maldacena}, it is known \cite{ryu,ryu2} that the entanglement entropy associated with a spatial subregion $A$ on the boundary is given by the area of a minimal surface in the bulk anchored on the boundary of $A$.

Integral geometry offers nice mathematical tools in building up the bulk
space.  The Crofton's formula expresses geometric objects of various dimensions
in terms of some reference objects that have clearer field theory correspondence,
like minimal surfaces \cite{santal,diffentropy,Czech:2014ppa,czech1}. Moreover, the OPE blocks \ie irreducible
representations in OPE are identified with local operators in the kinematic
space, which can in turn be obtained via Radon transform of local operators
in the AdS space \cite{Lin:2014hva,wittendiagram,czech2,deBoer:2016pqk}.
 
In this note we will discuss the subtle question about the kinematic space (a set consisting of the reference objects 
aforementioned) associated with a certain subregion. For simplicity we only consider Lorentzian AdS$_3$ (LAdS$_3$) and the subregion $A$ is chosen as an interval on a certain time slice (\eg $t=0$). There is a space-like
region $V_A$ bounded by $A$ and its Ryu-Takayanagi surface $\xg_A$. According to the subregion-subregion duality \cite{Jafferis:2015del,Dong:2016eik}, we are supposed to recover from the reduced density matrix $\rho_A$ anything within the entanglement wedge $\CW_A$, which is the causal development of $V_A$. To spot the issue, let us first narrow down to $V_A$. Some geodesics cross the RT surface $\xg_A$ and have portions lying outside $V_A$ (see \eg fig.~\ref{KAfig}(a)), which becomes an obstacle in the construction of the kinematic space. The reason is as follows. The measure of the kinematic space is given by the second derivative of the length of a geodesic, the latter of which corresponds to the entanglement entropy of some interval. However in the current case, one end of the interval lies outside $A$ and hence we are not supposed to know the entanglement entropy from $\rho_A$. We will also have the same issue for geodesics lying only partially in $\CW_A$. In other words, subregion-subregion duality seems to forbid the construction
of the kinematic space of a subregion. 

One solution is to simply drop the part outside $\CW_A$ \cite{Espindola:2018ozt}. It is known \cite{santal,huang} that
the correct measure can be obtained using end points on essentially arbitrary surfaces (in the current case one end on the
boundary and the other on $\xg_A$). On the field theory side, $\rho_A$ can be purified \cite{terhal2002entanglement,Takayanagi:2017knl,Nguyen:2017yqw} by an excited state that corresponds to the closed surface $\xg_A \cup A$ in the sense of surface/state correspondence \cite{Miyaji:2015yva}. The length of such a geodesic then has the interpretation as entanglement entropy of subsystem $A_1 \cup A_1'$ (with $A_1 \subset A$, $A_1' \subset \xg_A$), which is a generalized version of the entanglement of purification. The remaining problem is that it is difficult to describe the purification state involving DoFs living on the RT surface. We only know it has no spatial entanglement for the subsystem on $\xg_A$. There is no concrete field theory realization of the length of the geodesic discussed above.    

Recently, it was proposed \cite{Dutta:2019gen} that reflected entropy can provide an alternate interpretation of the entanglement wedge cross section
$E_W$ (see also \cite{Tamaoka:2018ned} for yet another interpretation of $E_W$). To define reflected entropy, one performs some sort of doubling trick to a reduced density matrix. More precisely the bras in the dual space are turned into kets and the mixed state then becomes a pure state in the extended Hilbert space, which can be viewed as the canonical purification of a density matrix. The structure of the dual Hilbert space is the same as the original space and hence a subsystem $A_1$ will have a natural image
of $A_1'$ in the dual Hilbert space. The entanglement entropy associated with the combined
subsystem $A_1 \cup A_1'$ then gives the reflected entropy. The doubling also leads to a bulk region $V_{A'}$ as the image (attached to $V_A$ by gluing together $\xg_A$ and its image) and the holographic dual of the reflected entropy is the minimal geodesic starting from the boundary of $A_1$ and ending on the boundary of $A_1'$, which is essentially precisely twice of $E_W$.

The canonical purification and the image $V_{A'}$ allows natural extension of the geodesics crossing $\xg_A$. For our purpose we need to consider the case when $A_1'$ is generic and not necessarily the image to $A_1$. We believe that the dual in the the bulk is a ``reflected geodesic'' \footnote{The term ``reflected geodesic'' was introduced in \cite{Dutta:2019gen} only for the case in which $A_1'$ is the image of $A_1$. Here we abuse it for more general configurations.}. More precisely $A_1$ and $A_1'$ are specified by two points $P,Q$ on the boundary, inside the region $A$ and its image $A'$ respectively. There
will be a unique point $X$ on the RT surface that minimizes the total length of two pieces of geodesics starting from $P$ and $Q$, the union of which
is what we call a reflected geodesic. Now we can construct the kinematic space using the second derivative of the lengths of the reflected geodesics. It is not difficult to see that the kinematic space constructed this way can be identified with the kinematic space that consists of geodesics crossing $\xg_A$. More importantly, we compute the length of a reflected geodesic using the correlator of twist operators $\Tr_A [\rho_A^{1/2} \xs(q) \rho_A^{1/2} \tilde \xs(p)]$, which only depends on the reduced density matrix $\rho_A$. The correlator is worked out using the replica trick \ie $\lim_{m\to 1/2} \Tr_A [\rho_A^{m} \xs(q) \rho_A^{m} \tilde \xs(p)]$. The two operators are separated by $m$ copies and a phase shift $q \to e^{2 \pi i} q$ takes a point to the next sheet. So $m=1/2$ then implies the correlator follows from $\br \xs(e^{\pi i}q)\tilde \xs(p) \ke$.

We also find that the reflection is in fact realized via some discrete symmetry,
which becomes crucial in extending the construction to the whole entanglement wedge. In general a reflected geodesic is obtained by applying the
discrete symmetry on the portion outside the entanglement wedge. The side
effect is that the reflected geodesic becomes discontinuous but the total length can still be
obtained via the correlator of twist operators.

The outline of the rest of the paper is as follows. In sec.~\ref{sec:kinspace}, we will give a brief review explaining what kinematic space is and why it is useful. In sec.~\ref{sec:reflectedg} we analyze the main issue and show that it can be resolved by considering
the kinematic space of the reflected geodesics, the latter of which is shown
to be equivalent to a subspace of the original kinematic space. We also compute
the length of the reflected geodesic entirely from the field theory side
using the correlator $\Tr_A [\rho_A^{1/2} \xs(q) \rho_A^{1/2} \tilde \xs(p)]$. In sec.~\ref{sec:causalwedge}, we generalize the construction
to the whole entanglement wedge. Finally we conclude in sec.~\ref{sec:conclusion}
and discuss some possible future directions.
 
\section{Kinematic space}
\label{sec:kinspace}

So let us quickly go through some important facts about integral geometry. Such a mathematical branch is not very well known in the high energy community, at least not until recently. One of the more important applications is the Crofton's formula, which expresses a geometric object in the real space as an integral over some reference objects. For our purpose the reference object is always chosen as geodesic. Let us first consider a static slice of the Lorentzian AdS$_3$. Henceforth we will work in the Poincare patch in which the metric reads
\begin{equation}\label{metric}
   \rmd s^2=\frac{\rmd z^2+\rmd x^+\rmd x^-}{z^2}=\frac{\rmd z^2+\rmd x^2-\rmd t^2}{z^2},
\end{equation}
where for later convenience we introduce the light-cone coordinates
\begin{equation}
    x^+=x+t\qquad x^-=x-t\,.
\end{equation}
 
The Crofton's formula in the two dimensional hyperbolic space $\BH^2$ (called real space) then reads
\be
\label{croftonH2}
  L_{M_1} =\frac{1}{2}\int_{M_1\cap \xg\neq0} N(M_1\cap \xg)\frac{\partial^2L_\xg(u,v)}{\partial u\partial v}\rmd u\wedge \rmd v.
\ee
Then Crofton's formula tells us that the length of an one dimensional curve $M_1$ in the bulk is given by the number of geodesics $\xg$ it has intersection with ($N(M_1\cap \xg)$ is the intersection number providing a weight if
$\xg$ hits $M_1$ multiple times). Of course, to do the integral, we need a measure. The collection of all the geodesics forms a space called the kinematic space in which every point represents a geodesic. Generically, we use the end points $u,v$ to parameterize a geodesic. 
The measure can be obtained by the second derivative of the length of a geodesic 
\be\label{kinmeasurepoin}
    \rmd s^2=2\frac{\partial^2L_\xg(u,v)}{\partial u\partial v} \rmd u\rmd v=\frac{4\rmd u\rmd v}{(u-v)^2}\,,
\ee
where the length of a geodesic is given by $L_\xg(u,v)=\log[(u-v)^2/\xe^2]$.
Now this expression has a natural entropic interpretation since the length is related to the entanglement entropy of an interval of $[u,v]$ via Bekenstein-Hawking formula $S(u,v) = L_\xg(u,v)/4 G$ ($G= 3 L/2 c$ from AdS$_3$/CFT$_2$ with
$L$ being the AdS scale and $c$ the central charge). The point is that this quantity can be computed from the field theory side and hence we can determine the geometry of the kinematic space using the entanglement entropy from the boundary field theory. Moreover every point in the real space can be realized as a geodesic (so-called point curve) in the kinematic space and the distance between any two points can be be expressed as a volume integral \cite{Czech:2014ppa,czech1}. 

So far the real space is one $\BH^2$ time slice. We can go one step further
and consider geodesics in the whole AdS$_3$ space. Only space-like geodesics
will be included since it is less clear what corresponds to time-like or null geodesic in the field theory. The geometry of this ``covariant'' kinematic
space is again given by the second derivative of the length in the sense that 
\be \label{measurederi}g_{\mu\nu} =\p_\mu \p_\nu L_{\xg}\,,\ee 
which gives in the current case
\be
\label{3adskin}
\rmd s^2=\frac{2\rmd u^+\rmd v^+}{(u^+-v^+)^2}+\frac{2\rmd u^-\rmd v^-}{(u^--v^-)^2}.
\ee
The length $L_\xg (u^\pm ,v^\pm)$ of a geodesic with end points $u^\pm ,v^\pm$ in the light-cone coordinates reads ($\xe$ being the cutoff)
\begin{equation}\label{3adsgeol}
  L_\xg (u^\pm ,v^\pm)=\log\lsb\frac {(u^+-v^+)({u^-}-{v^-})} {\xe^2}\rsb\,.
\end{equation}
In fact, the formula of second derivative works in a general space, which is the starting point of our later construction.

\section{Kinematic space associated with an interval}
\label{sec:reflectedg}
Let us now get down to the entanglement wedge associated with a subregion $A$, or more precisely the subspace of $\BH^2$ (denoted by $V_A$) that is bounded by $\xg_A \cup A$ ($\p V_A = \xg_A \cup A$). For simplicity, we only consider the case when $A$ is a single interval. There is the so-called subregion-subregion duality \cite{Jafferis:2015del,Dong:2016eik} in AdS/CFT. The original formulation is in terms of relative entropy. Basically it says the same reduced density matrix in the boundary theory leads to the same reduced density matrix in the bulk associated with the subregion $V_A$. Such a duality implies that the information of a subregion on the boundary is good enough to construct a subregion in the bulk. 

\begin{figure}[tbp]
\begin{center}
\begin{minipage}[t]{0.49\textwidth}
\includegraphics[scale=0.5]{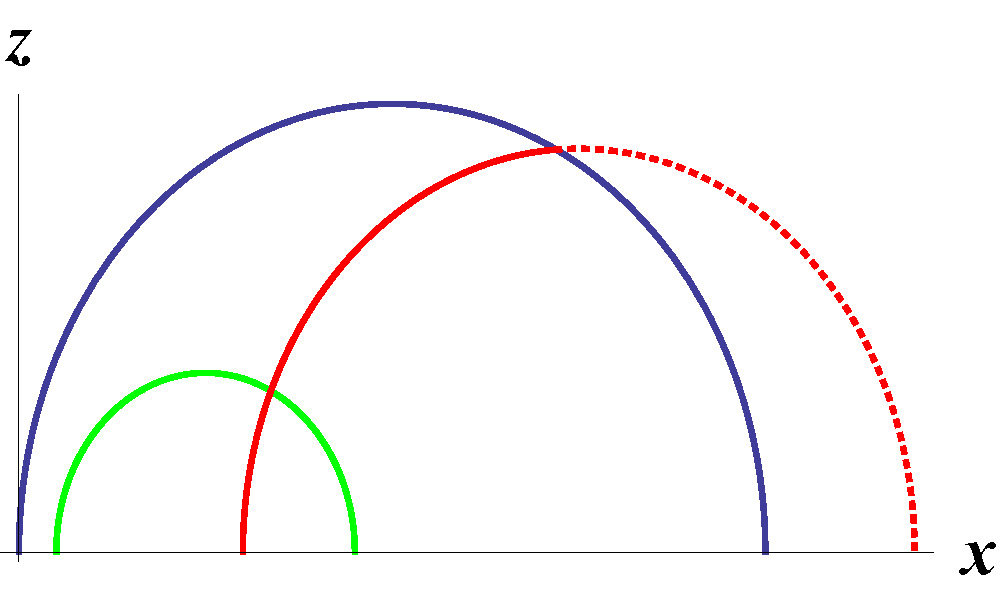}
\end{minipage}
\begin{minipage}[t]{0.49\textwidth}
\includegraphics[scale=0.5]{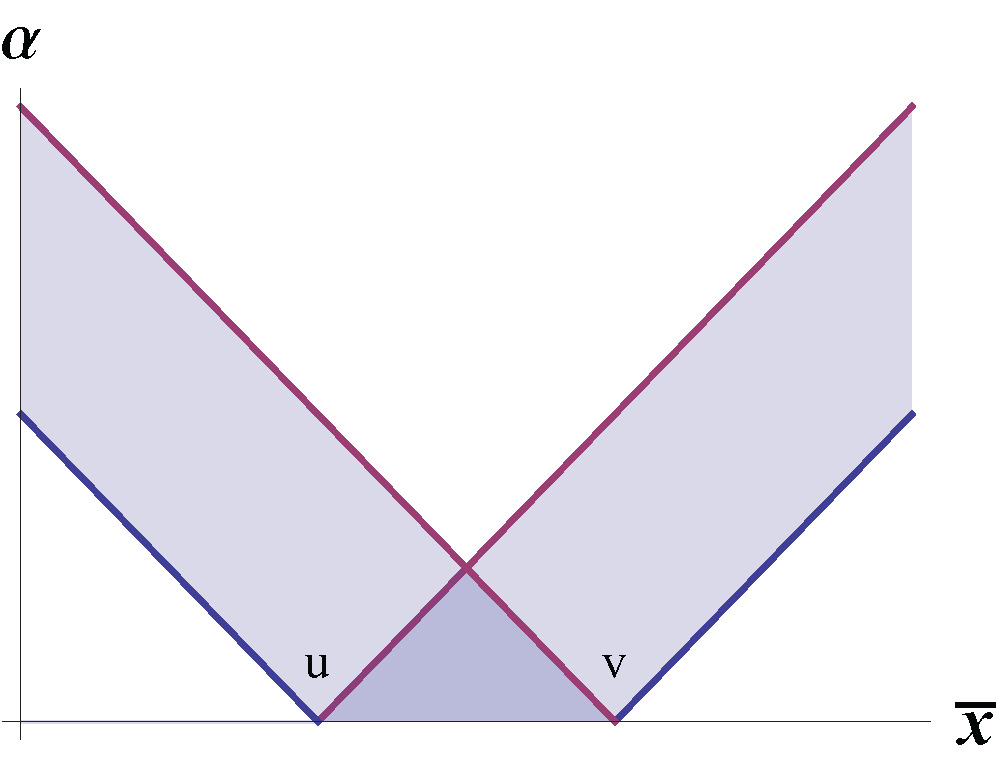}
\end{minipage}
\caption{(a) (left) There are two types of geodesics in $V_A$: I. both end points on $A$ (green); II. crossing $\xg_A$ (red). (b) (right) Total kinematic space $\CK$ in coordinates $\bar x = \frac {u+v} 2, \xa = \frac {v-u} 2$. The kinematic space $\CK_{V_A}$ associated with $A$ is painted with the shadow. Points in the darker triangle denote those with both end points on $A$ (type I) while points in the two rectangles describe geodesics crossing $\xg_A$ (type II).}
\label{KAfig}
\end{center}
\end{figure}

The subregion-subregion duality requires that the length
of any bulk curve within the entanglement wedge shall be expressed in terms of the reference
objects entirely within the subregion. Unfortunately, the Crofton's formula given above is not up to the task. Not all the geodesics are lying entirely inside
the subregion $V_A$ (see fig.~\ref{KAfig}(a)) but we need all their lengths $L_\xg$ to define the measure. Clearly we need a different construction for the geodesics crossing the RT-surface.

\subsection{Purification and reflected entropy}
From the geometric stand point, the resolution is to simply cut off the part
of a geodesic outside $V_A$. The fact that \cite{santal,huang} the definition by eq.\er{measurederi} always works \footnote{One can choose a different surface $\xS'$ as the boundary and the measure as a two form satisfies $\p_u\p_v L_\xg(u,v) \rmd u \wedge \rmd v = \p_{u'}\p_{v'} {\tilde L}_\xg(u',v') \rmd u' \wedge \rmd v'$, where $u',v'$ are new end-point coordinates on $\xS'$ and $\tilde L$ is the corresponding length.} implies that some suitable coordinate $x_0$ for the end point $X$ on the RT surface should be enough for constructing the kinematic space. More precisely, the
kinematic space $\CK_{V_A}$ for $V_A$ is a subspace (see fig.~\ref{KAfig}(b)) of the total kinematic space $\CK$ (given by eq.\er{kinmeasurepoin}) and $\p_x \p_p\ell(\xg_{PX})$ gives the same measure (up to a coordinate transformation from $q$ to $x$). $\ell(\xg_{PX})$ is the length of the geodesic $\xg_{PX}$. Here we use the notation $\xg_{AB}$ to denote a geodesic from $A$ to $B$, both of which can be either in the bulk or on the boundary.

\begin{figure}[tbp]
\begin{center}
\includegraphics[scale=0.7]{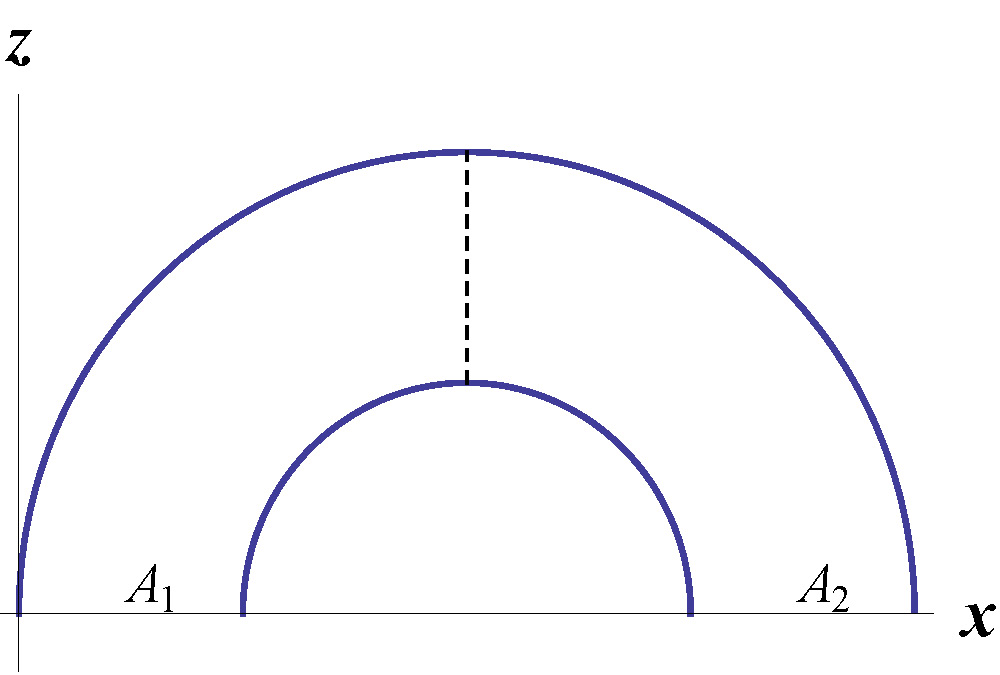}
\caption{$A = A_1 \cup A_2$ consists of two disjoined regions $A_1$ and $A_2$. The blue semi-circles are the RT surface $\xg_A$ and the dashed line gives the entanglement wedge cross section.}
\label{EWfig}
\end{center}
\end{figure}

Subregion-subregion duality requires that $\ell(\xg_{PX})$ comes from the data in $A$. Indeed such a candidate can be found with the help of purification \cite{terhal2002entanglement,Takayanagi:2017knl,Nguyen:2017yqw}. In general the purification of a system $A$ can be understood as a pure state $ \st{\psi_{AA'}}$ that lives in the tensor product of $\CH_A \otimes \CH_{A'}$ and gives the right reduced density when $\CH_{A'}$ is traced out,
\[\Tr_{A'} \st{\psi_{AA'}}\br\psi_{AA'}| = \rho_A\,.\] 
To reproduce a certain reduced density matrix $\rho_A$, there are infinite number of ways to do the purification. If $A$ consists of two subsystems $A_1, A_2$, among all possible purifications and partitions of $A_1',A_2'$ there is one minimizing the entanglement entropy of $S(\rho_{A_1 \cup A_1'})$, where $A_1'$ is a subsystem of the
purification system ($A = A_1 \cup A_2$, $A' = A'_1 \cup A'_2$). This is by definition the entanglement of purification $E_P$
\be
E_P(A_1:A_2) = \zt{min}\, S(\rho_{A_1 \cup A_1'}) = E_W(A_1:A_2) 
\ee
and its holographic dual (if available) is proposed to be the entanglement wedge cross section $E_W$, which is the minimal surface stretching  between the RT-surfaces associated with $A_1 \cup A_2$ (see fig.~\ref{EWfig}).

\begin{figure}[tbp]
\begin{center}
\includegraphics[scale=0.7]{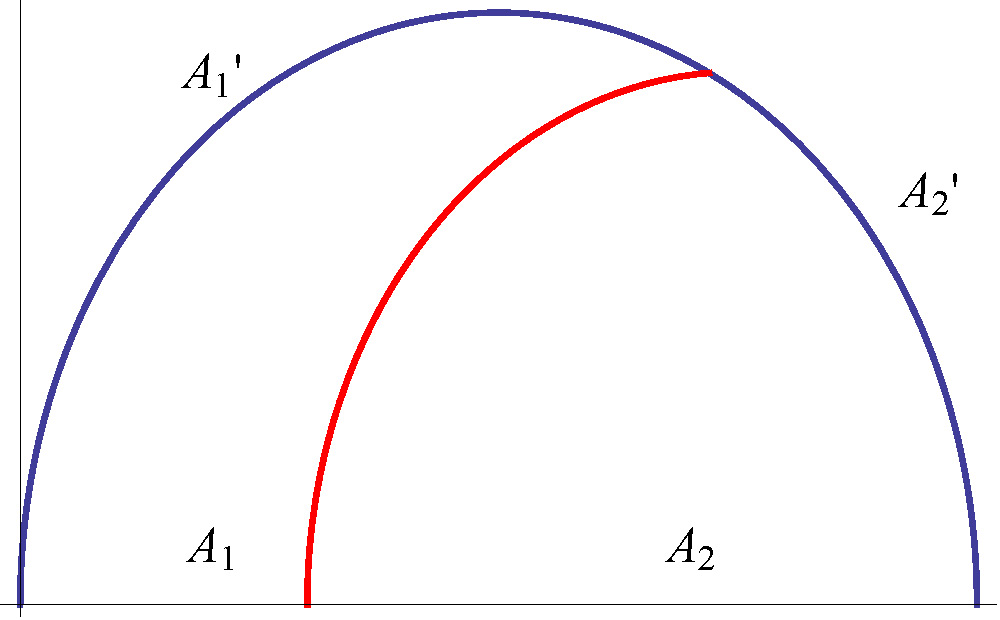}
\caption{$A = A_1 \cup A_2$ consists of two regions $A_1$ and $A_2$ sharing the same boundary. With appropriate choice of $A_1'$, $S(\rho_{A_1 \cup A_1'})$ is realized as a geodesic (red) ending on $\xg_A$.}
\label{geopurification}
\end{center}
\end{figure}

Heuristically one can think that the optimized purification is realized by a subsystem $A' = A_1' \cup A_2'$ living on the RT surface $\xg_A$ (associated with the subregion $A$), which is obtained by applying RG flow only on the subregion $\overline{A} = \overline {A_1 \cup A_2}$ \footnote{The readers shall beware that it is far from clear whether the optimized purification is realized by a subsystem on the RT surface. Nor is there any rigorous proof that $E_P = E_W$.}. We
note that surface-state correspondence \cite{Miyaji:2015yva} implies that the closed surface $\xg_A \cup A$ corresponds to a pure state the vacuum
can flow into via entanglement renormalization. The optimized total wave function $\psi_{A_1A_2A_1'A_2'}$ should have no extra entanglement for the DoFs in $A_1' \cup A_2'$. For our interest, $A$ is taken to be a single interval (\ie $A_1 \cap A_2 \ne \emptyset$) and for the optimized purification $\psi_{A_1A_2A_1'A_2'}$, a non-optimized partition of $A'$ can be used to define the entanglement entropy $S(\rho_{A_1 \cup A_1'})$, which is expected to correspond to a geodesic ending on the RT surface \cite{Espindola:2018ozt} (see fig.~\ref{geopurification}). As explained earlier the kinematic space can be constructed using the lengths of these geodesics and now we can see that they indeed have information theoretic interpretation on the field theory side.

However, this is not the whole story since practically there is little way to figure out the purification $\psi_{A_1A_2A_1'A_2'}$, which makes the field theory computation of $S(\rho_{A_1 \cup A_1'})$ very difficult. The problem can be better handled if the purification is restricted to a special class (see \eg \cite{Hirai:2018jwy} \cite{Caputa:2018xuf}). 
 
\subsection{Reflected geodesic}
\label{subsec:reflctedg}
Recently it was proposed that the notion of reflected entropy provides an alternate interpretation of $E_W$ \cite{Dutta:2019gen}, which as we shall see is enough to build the kinematic space from the information theoretic quantities in the field theory. To define reflected entropy it is necessary to first turn the density matrix $\rho_A$ (Here $A = A_1 \cup A_2$ is again a certain spatial subregion) into a pure state, \ie the canonical purification by switching bras into kets
\[\st{\sqrt{\rho_{A}}} \in \zt{End}\,{\CH_{A}} = \CH_{A} \otimes \CH^*_{A}\,.\]
In other words, the Hilbert space associated with $A$ is doubled 
\[\CH_A \to \CH_A \otimes \CH_{A'},\quad \CH_{A'} \cong \CH^*_{A}\,.\]
The pure state $\st{\sqrt{\rho_A}}$ lies in the extended Hilbert space and gives back the reduced density matrix $\rho_A=A_{ii}{} \st{i}_{A}{}_A\br{i}|$
when $A'$ is traced out. The square root in $\st{\sqrt{\rho_{A}}}=A_{ii}{}^\h \st{i}_A\st{i}_{A'}$ is required to get the right density matrix
\[\rho_A =\Tr_{A'} \st{\sqrt{\rho_{A}}}\br
\sqrt{\rho_{A}}|=\Tr_{A'}\lb A_{ii}{}^\h \st{i}_{A}\st{i}_{A'}A_{jj}{}^\h {}_{A}\br j|{}_{A'}\br j|\rb=A_{ii} \st{i}_{A}{}_{A}\br{i}|\,.\]

As $A'$ is essentially the image of $A$, we can choose the same decomposition $A' = A_1' \cup A_2'$ so that $A_1'$ and $A_2'$ are the images of $A_1$
and $A_2$. In this scenario, the reflected entropy $S_R(A_1:A_2)$ is defined as the entanglement entropy $S(A_1 \cup A_1')_{\st{\sqrt{\rho_A}}}$ of subsystem $A_1 \cup A_1'$, which is shown to be twice the entanglement of purification
\be\label{reflectede} S_R(A_1:A_2) \equiv S(A_1 \cup A_1')_{\st{\sqrt{\rho_A}}}= 2E_W(A_1:A_2)\,.\ee

\begin{figure}[tbp]
\begin{center}
\includegraphics[scale=0.8]{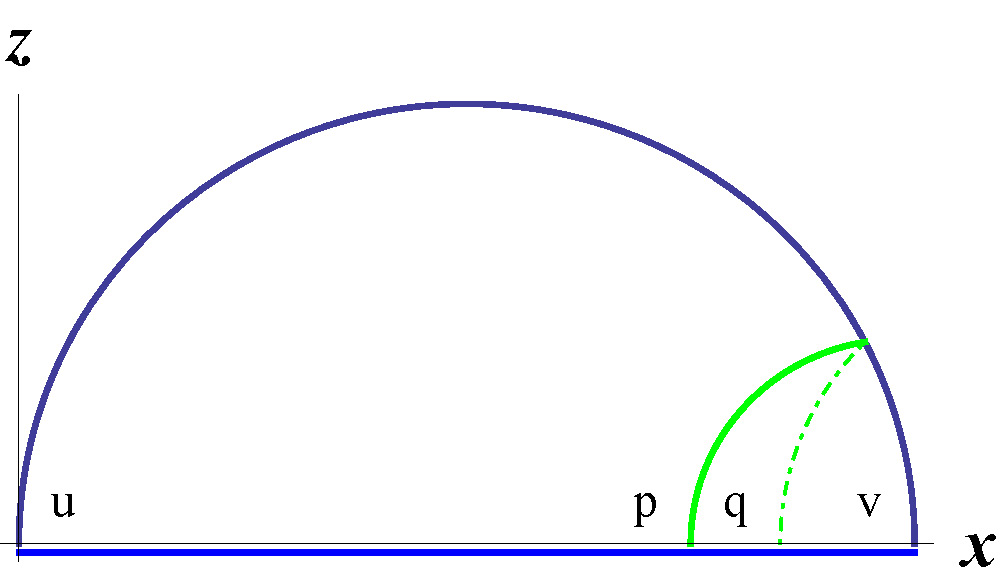}
\caption{Reflected geodesic for $A_1 = [u,p]$ and $A_1' = [u,q]$. We overlay $V_A$ with its image $V_{A'}$ even though they are two different bulk subregions. They indeed share the same RT surface. The solid green line denotes $\xg_{PX}$ while the dashed green line denotes $\xg_{QX}$ lying in $V_{A'}$.}
\label{reflectedgeo}
\end{center}
\end{figure}

According to the subregion-subregion duality (if AdS/CFT is applicable), each copy of $A$ shall lead to
a bulk subregion. The bulk (real space) then consists of a subregion $V_A$ and its image $V_{A'}$ glued together via the RT surface. As in the discussion of purification, there is nothing to stop us from decomposing
the $A'$ into $A_1' \cup A_2'$ differently so that $A_1'$ is no longer
isomorphic to $A_1$. In any event, the entanglement entropy $S(A_1 \cup A_1')_{\st{\sqrt{\rho_{A}}}}$
is expected to be dual to some geometric object, which as we shall see consists of two piecewise smooth geodesics glued together somewhere on
the RT surface $\xg_A$. This is what we call a reflected geodesic. It can be understood as a geodesic that crosses $\xg_A$ into the image $V_{A'}$ and eventually lands on $A'$. From our previous experience, the kinematic space is usually insensitive to how the geodesics is extended. In other words, canonical
purification shall work just as well as the optimized purification. 

For computational purpose, we can think of the reflected geodesic bouncing off $\xg_A$ back to $A$ (see fig.~\ref{reflectedgeo}). The transition from $V_A$ to $V_{A'}$ is not necessarily smooth. In fact, we propose such a rule that the total action (\ie length) needs to be minimized, which is the same condition for a reflection. We therefore use the name reflected geodesic for such a pair of geodesics.     

For a single interval $A=[u,v]$, $A_1 = [u,p],A_1' = [u,q]$, the entanglement entropy $S(A_1 \cup A_1')_{\st{\sqrt{\rho_{A}}}}$ shall match the length a reflected geodesics ($X$ picked to make the total length minimized)
\[S(A_1 \cup A_1')_{\st{\sqrt{\rho_{A}}}} = \ell(p,X)+\ell(X,q)\,,\]
We first work out the total length on the right hand side. The length of each arc of the reflected geodesic is derived in appendix~\ref{lengthgeo} and the sum reads
\[\ell(p,x_0) + \ell(q,x_0)= \log \left[\frac{b^2+p (p-2 x_0)}{\sqrt{b^2-x_0^2}}\right]+\log \left[\frac{b^2+q (q-2 x_0)}{\sqrt{b^2-x_0^2}}\right]\,.\]
where $x_0$ is the $x$-component of the coordinates of the point $X$. Extremization of $\ell(p,x_0) + \ell(q,x_0)$ leads to the following solutions
\be
\label{xsol1}
x_{0;1}(p,q)= \frac{p (2 q-u-v)-q (u+v)+u^2+v^2}{2 (p+q-u-v)}\,,\ee
or
\be
\label{xsol2}
x_{0;2}(p,q) = \frac{(u-v)^2 (p+q-u-v)}{2 \left[p (2 q-u-v)-q (u+v)+u^2+v^2\right]}\,.
\ee
The first solution is valid when one of the points
$P,Q$ is outside the interval. On the other hand, the second solution gives the length of the reflected geodesic when both are inside
\ba
\label{lreflectedgeo}
 \tilde S(p,q)&= & \log \left[\frac {(\frac{v - u}2)^2-(p-\frac{v + u}2) (q-\frac{v + u}2)}{(\frac{v - u}2)}\right]^2\nn
& = & \log \left[\frac{p (-2 q+u+v)+q (u+v)-2 u v}{u-v}\right]^2\,,
\ea
where we use $\tilde S$ to denote the length as its functional form is different from $S(p,q) = \log (p-q)^2$.

For convenience, we turn $\tilde S$ into the following form after a subtraction of the length of a minimal geodesic $\log (p-q)^2$ (which as we shall see in sec.~\ref{sec:fieldtheorycom} is how the conformal block is extracted from a 4-point function by removing a 2-point function) 
\be \CF(z^+) = \tilde S(p,q)-\log (p-q)^2=2\log \frac {{z^+}+1}{{z^+}-1}\label{reflectedez}\ee
that depends only on the cross-ratio
\be\label{crossratio} {z^+} = \frac {(p-u)(q-v)}{(q-u)(p-v)}\,.\ee
We will later recover this form $\CF$ from the CFT computation. Before we move on to the kinematic space, we would like to comment on the other solution. It turns out that the two are related by a discrete symmetry.
For simplicity, we use the translational symmetry to set $v=-u=b$. The discrete
symmetry is then given by 
\be I: x \to \frac {b^2} x\,. \ee
More precisely, we have 
\[x_{0;2}(p,q) = x_{0;1}(p,q'),\quad q' = b^2/q\,.\]
The symmetry can be extended to the bulk
\be\label{discreteIct} I:\quad z \to \frac {b^2z} {x^2+z^2}\,, \quad x \to \frac {b^2x} {x^2+z^2}\,.\ee
One can easily see that a point $X$ on the RT surface is invariant under $I$. In other words, $I$ takes the geodesic $\xg_{QX}$ to $\xg_{Q'X}$. Moreover, this is a conformal transformation on the boundary (and isometry in the bulk) as it is the inversion $x^\mu \to x^\mu/x^2$ sandwiched
by the scaling transformation $x^\mu \to \iv b x^\mu$ and its inverse.

\subsection{Measure}
The geodesics in the total kinematic space can be divided into three classes: a) $P,Q \in A$; b) $P,Q \in A'$; c) only one of $P,Q \in A$ (for the unoriented geodesics we are going to work with, $P$ is always chosen to be in $A$). Those geodesics in the b class are irrelevant for the geometry in $V_A$ as their intersection numbers always vanish. The kinematic space in the a class is trivial and is a subspace of the total kinematic space (the triangular region in fig.~\ref{KAfig}(b)). So we are left with those in the c class, which as we will see can be equivalently described by the reflected geodesics.

The measure in the kinematic space is given by the second derivative of the length of a geodesic and such a definition is insensitive to
the end points of the geodesic in the sense that \cite{santal,huang} we can choose essentially arbitrary cutoff surfaces to define end points (which leads to different lengths). It is less obvious but somewhat expected that the formula shall remain valid even if the geodesics hitting the RT
surface $\xg_A$ are extended to region $V_{A'}$. The extension is unique and provides a one-to-one correspondence between a geodesic hitting $\xg_A$ and a reflected geodesic. We therefore expect the kinematic space of the reflected geodesics shall agree with that of the class c geodesics (cut off at $\xg_A$). 

The proof only costs a few lines of algebra. Assuming a
geodesic is labeled by its end point coordinates $(p,x_0)$, the second derivative
then becomes $\p_p \p_x \ell(p,x_0)$. On the other hand, the second derivative of
the reflected geodesic is given by 
\be\p_p \p_q \tilde S(p,q) =\p_q \lsb \p_p \ell(p,x_0)+ \p_{x_0} \ell(p,x_0)\frac {\p x_0}{\p p}\at{q}+\p_{x_0} \ell(q, x_0) \frac {\p x_0}{\p p}\at{q}\rsb = \p_{x_0} \p_p \ell(p,x_0)\frac {\p x_0}{\p q}\at{p} \ee
In the second equality, we use the fact that $\ell(p,x_0)+\ell(q,x_0)$ is minimized
with respect to the variation of $x_0$. The final form provides the same measure
$\p_p \p_{x_0} \ell(p,x_0)$ in a different coordinate $p,q$ with $\frac {\p x_0}{\p q}\at{p}$ being the Jacobian from $(p,x_0)\to(p,q)$. As explained earlier, taking the end point coordinate on $\xg_A$ shall give the same measure (up to a coordinate transformation). We then get to the conclusion that the kinematic space including the reflected geodesics is the same as $\CK_{V_A}$ with a different parameterization. 

For later convenience, we present a different derivation based on the discrete
symmetry $I$, which establishes the coordinate transformation $q \to q'$. The
piece $\xg_{QX}$ is mapped to $\xg_{Q'X}$ under $I$. However,
the subtlety is that their lengths are not equal as the cutoff in $z$ coordinate
does not respect this symmetry. Fortunately, the difference is $p$-independent
and hence drops out after a second derivative 
\[\tilde S(p,q) =2\log \lb b-\frac {p q}{b}\rb = 2\log \lb \frac {b^2} q-p\rb + 2\log \frac {q}{b} = \log (q'-p)^2 + 2\log \frac {q}{b}\,,\]
where we again use the translational symmetry to set $v=-u=b$. As a result, the measure is essentially $\p_p \p_q \ell(p,q') = \p_p \p_{q'} \ell(p,q') \frac {\p q'}{\p q} $, which is the measure in $\CK_{V_A}$ with different
coordinates $(p,q)$.  
  
\subsection{Field theory computation}
\label{sec:fieldtheorycom}
As pointed out earlier, constructing the kinematic
space itself is never a problem. What is really needed is the field theory definition of the measure, which we will obtain with the help of the concepts of reflected geodesic and reflected entropy. Our goal is to interpret $S(A_1 \cup A_1')_{\st{\sqrt{\rho_{A}}}}$
(which we call generalized reflected entropy) as the length of a reflected geodesic. As an entanglement entropy we expect that it can be computed from the 2-point
function of the twist operators, or equivalently the Virasoro vacuum OPE block of a pair of scalar operators $\CO$ (see \eg \cite{Fitzpatrick:2016mtp}). In fact similar setup can be found in an AdS black hole where $A_1$ and $A_1'$ are spatial regions on different
boundaries (see \eg \cite{Hartman:2013qma}). It is also shown \cite{Dutta:2019gen} that the generalized reflected entropy $S(A_1 \cup A_1')_{\st{\sqrt{\rho_{A}}}}$ can be computed using a 4-point function (obtained by inserting twist operators at $u,v$) that shall only depend on the cross ratio \er{crossratio}. Such observation greatly simplifies the computation. 

We find it easier to do this computation using the techniques developed in \cite{Faulkner:2018faa} for computing $\br \CO(q) \xD_A^{is} \CO(p) \ke$ ($\xD_A$ being the modular operator, see appendix~\ref{connectionmodular}), which itself can serve as an alternate dual for the length of a reflected geodesic (see appendix~\ref{connectionmodular} for the connection). More precisely, we consider its cousin with clearer entropic interpretation   
\be \lim_{n\to 1}\Tr_A [\rho_A^{n/2} \CO(q) \rho_A^{n/2} \CO(p)] = \br \CO(q)\xD^{\frac 1 2} \CO(p) \ke\ee
The quantity on the left hand side (introduced in \cite{Faulkner:2018faa} for computation purpose) is precisely what we need. It is essentially the 2-point function with respect to the state $\st{\sqrt{\rho_A}}$ with one operator $\CO(q)$ acting on the dual Hilbert space while the other $\CO(p)$ on the original Hilbert space
(we put a prime on $\CO(q)$ to emphasize that it is defined in the space $\CH_{A'}$).  
\[\br\sqrt{\rho_{A}}| \CO'(q) \CO(p)\st{\sqrt{\rho_{A}}}\,.\]
So the connection with the reflected entropy can be established.

\begin{figure}[tbp]
\begin{center}
\includegraphics[scale=0.5]{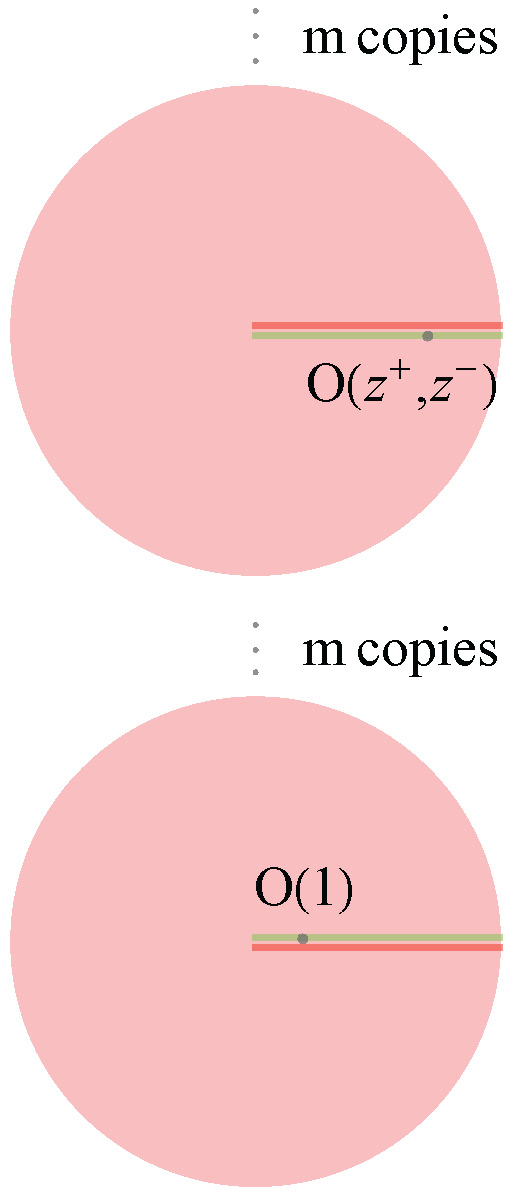}
\caption{Replica trick to compute $\Tr_A [\rho_A^{m} \CO(q) \rho_A^{m} \CO(p)]$ in terms of a path integral on the covering space. The two operators are separated by $m$ copies.}
\label{replica}
\end{center}
\end{figure}

To perform the computation we can first take $n = 2m$, then in the path integral language,
the 2-point function is computed on the $2m$-covering space with two operators
separated by $m$ copies (see fig.~\ref{replica}). We note that to move one operator to the next copy,
a phase factor $e^{2\pi i}$ is needed \footnote{This kind of analytic continuation is widely used in the computation of the entanglement entropy of excited states \cite{Asplund:2014coa} \ie the Virasoro block of HHLL operators \cite{Fitzpatrick:2015zha}. It is also similar to the use of KMS relation in thermal correlator.} and the 2-point function becomes
\be \lim_{m\to \h}\Tr_A [\rho_A^{m}\CO(q) \rho_A^{m}  \CO(p)] = \lim_{m\to \h} \br \CO(e^{2 m \pi i} q) \CO(p) \ke\,. \ee
Strictly speaking, this move only applies to the correlators in an Euclidean CFT$_2$ but the Lorentzian correlator can be obtained by analytic continuation from an Euclidean one. 

To achieve our goal of computing $S(A_1 \cup A_1')_{\st{\sqrt{\rho_{A}}}}$, we can take $\CO(q),\CO(p)$ to be twist operators $\xs,\tilde \xs$. Recall that the entanglement entropy (for a certain state $\st{\Psi}$) is given by the 2-point function of twist operators
\be S_{\zt{EE}} = \lim_{k \to 1}\frac 1 {1-k} \log \br\Psi| \xs (q)\tilde \xs (p)\st{\Psi} \ee
and the vacuum 2-point function of $\xs(q),\tilde \xs(p)$ goes like $|q-p|^{-\frac c {6} (k - \frac 1 k)}$. Now in the case of $m=1/2$, the phase shift $e^{2m\pi i}$ gives a negative sign ($z^+ \to - z^+$) to the vacuum correlator $|z^+-1|^{-\frac c {6} (k - \frac 1 k)}$ and the 2-point function simply reads 
\be\lim_{k \to 1} \frac 1 {1-k} \log \Tr_A [\rho_A^{\h} \xs(q) \rho_A^{\h} \tilde \xs(p)] = \frac c {6} \log (z^++1)^2\ee
Here we use conformal symmetry to put $u,v$ at $0, \infty$ and $p$ at $1$ as we know the correlator is equivalent to a 4-point function. To isolate the conformal block from a 4-point function that only depends on the cross ratio, we need to remove the 2-point function with both operators on the same sheet, which corresponds to the length of a minimal geodesic (\ie $\frac c {6} \log (z^+-1)^2$ after the operation by $\lim_{k \to 1} \frac 1 {1-k} \log$). The agreement with \er{reflectedez} can be seen immediately. 

\section{Kinematic space of an entanglement wedge}
\label{sec:causalwedge}
Our next step is to extend the construction above to the entanglement wedge $\CW_A$, which is the causal
development of $V_A$. For simplicity, we assume $A$ is an interval on the $t=0$ slice. Generalization to the case with two end points $u,v$ having different times is straightforward. We will further use the translational symmetry to set $v=-u=b$.

So we would like to consider the case when $P,Q$ have different time coordinates. It is tempting to try the ansatz of two geodesics $\xg_{PX},\xg_{QX}$ meeting at point $X$ on the light-like boundary of the entanglement wedge. As in the previous case, we can apply variation principle to the total length
\[\ell(p_1, p_2;x_0,y_0)+\ell(q_1, q_2;x_0,y_0)\,,\]
each of which is given by \er{geodesicl2}. There will again be more than one solutions. However, it turns out that in this covariant scenario, only the following solution makes sense when either $P$ or $Q$ is outside the causal diamond associated with $A$
\ba
\label{solxy}
x_0(P,Q)& = &  \frac{b^2 (p_1-q_1)+2 b (p_2 q_1-p_1 q_2)-q_1 \left(p_1 (q_1-p_1)+p_2^2\right)+p_1 q_2^2}{2 b (p_2-q_2)+p_1^2-p_2^2-q_1^2+q_2^2}\,,\nn
y_0(P,Q)& = &  \frac{b^2 (p_2-q_2)+p_1^2 q_2-p_2 \left(q_2 (p_2-q_2)+q_1^2\right)}{2 b (p_2-q_2)+p_1^2-p_2^2-q_1^2+q_2^2}
\ea
This is the case when such a pair of space-like separated points determines a geodesic via variation principle. There is indeed another solution for the case when both are inside the causal diamond but it can be checked that the other solution does not reduce to \er{xsol2} when both $P,Q$ are on the same time slice.

This issue can be resolved utilizing the discrete symmetry obtained earlier in eq.\er{discreteIct}. Including the time coordinate, the discrete symmetry we need is
\be I:\quad t \to -\frac {b^2t} {-t^2+x^2},\quad x \to \frac {b^2x} {-t^2+x^2}\ee
It is the inversion (sandwiched by scaling) plus time reversal. One can see the
tips of the causal diamond are invariant. In AdS$_3$, it is extended to
\be I:\quad z \to \frac {b^2z} {-t^2+x^2+z^2}\,, \quad t \to -\frac {b^2t} {-t^2+x^2+z^2},\quad x \to \frac {b^2x} {-t^2+x^2+z^2}\ee
and one can show that it is an isometry of LAdS$_3$. We can play the same trick by applying the symmetry $I$ on the end point $Q$, which takes it out of the causal diamond
(the image denoted as $Q'$). The solution \er{solxy} then gives us a point on the light-like boundary of the entanglement wedge
\ba
\tilde x_0(P,Q) & = & x_0(P,Q') = \frac{b^2 \left[b^2 (p_1+q_1)-2 b (p_1 q_2+p_2 q_1)-q_1 \left(p_1 (p_1+q_1)-p_2^2\right)+p_1 q_2^2\right]}{b^4-2 b^3 q_2+2 b p_2 \left(q_2^2-q_1^2\right)-(p_1^2-p_2^2) (q_1^2-q_2^2) }\,,\nn
\tilde y_0(P,Q) & = & y_0(P,Q') =\frac{b^2 \left[b^2 (p_2-q_2)+p_1^2 q_2-p_2 \left(q_2 (p_2-q_2)+q_1^2\right)\right]}{b^4-2 b^3 q_2+2 b p_2 \left(q_2^2-q_1^2\right)-(p_1^2-p_2^2) (q_1^2-q_2^2)}\,.\label{coordinatesX}
\ea
The major difference compared to the points on the RT surface is that $X$ (whose coordinates are given by $(z,t,x) =(\sqrt{(b-\tilde y_0)^2-\tilde x_0^2},\tilde
y_0, \tilde x_0)$) is no longer invariant under $I$ even though the image $X'$ is still on the
light cone   
\[\left(\frac{b^2 \sqrt{(b-\tilde y_0)^2-\tilde x_0^2}}{(b-\tilde y_0)^2-\tilde y_0^2}\right)^2+\left(\frac{b^2 \tilde x_0}{(b-\tilde y_0)^2-\tilde y_0^2}\right)^2-\left(b+\frac{b^2 \tilde y_0}{(b-\tilde y_0)^2-\tilde y_0^2}\right)^2 = 0\]
Previously, we have seen that the geodesic $\xg_{Q' X}$ is related to $\xg_{Q X}$ by $I$. Despite the lack of invariance of $X$, which is fixed by the
image of $Q$ via \er{coordinatesX}, it is still tempting to
consider the total length of the two pieces of geodesics $\xg_{PX}$ and $\xg_{QX'}$. Both lengths can be computed using the formula \er{geodesicl2} and the sum is given by ($u,v$ are restored using translational symmetry)
\be
\tilde S(P,Q) = \log \frac {\left[(\frac{v - u}2)^2-(p_1-p_2-\frac{v + u}2) (q_1-q_2-\frac{v + u}2)\right] \left[(\frac{v - u}2)^2-(p_1+p_2-\frac{v + u}2) (q_1+q_2-\frac{v + u}2)\right]}{(\frac{v - u}2)^2}
\ee
In order to study a Lorentzian CFT$_2$, it is convenient to introduce light-cone
coordinates $x^+ = x+t, x^- = x - t$. The discrete symmetry is realized as
\be\label{inverselc} I: x^+ \to b^2/x^+,\quad x^- \to b^2/x^-\ee
We can then express $\tilde S(P,Q)$ in the
factorization form of a holomorphic and an anti-holomorphic part
\be
\tilde S(P,Q)=\log \left[\frac {(\frac{v^- - u^- }2)^2-({p^-}-\frac{v^- +
u^-}2) (q^--\frac{v^- + u^- }2)}{(\frac{v^- - u^- }2)}\right]\left[\frac
{(\frac{v^+ - u^+}2)^2-(p^+-\frac{v^+ + u^+}2) (q^+-\frac{v^+ + u^+}2)}{(\frac{v^+ - u^+}2)}\right]
\ee
We note that both end points $u,v$ of the interval have $t=0$ and hence satisfy
$u^-  = u^+, v^- = v^+$. It is easy to see that both factors take the same
form as in \er{lreflectedgeo} and hence can be expressed in terms of the cross
ratio (after a subtraction of $\log [(u^+-v^+)(u^--v^-)]$) 
\be \CF(z^+,z^-) =\log \lb\frac {z^+ +1}{z^+-1}\rb\lb\frac {z^-+1}{z^--1}\rb,\quad z^\pm = \frac {(p^\pm-u^\pm)(q^\pm-v^\pm)}{(q^\pm-u^\pm)(p^\pm-v^\pm)}\ee
The field theory computation in sec.~\ref{sec:fieldtheorycom} can be carried over to the case of $z^+ \ne z^-$ even though the entropic interpretation is
less clear for a general configuration of $p^\pm,q^\pm,u^\pm,v^\pm$. Because of the factorization, it is straightforward to see that $\CF(z^+,z^-)$ can be
reproduced from the correlator $\Tr_A [\rho_A^{1/2} \xs(q^+,q^-) \rho_A^{1/2} \tilde \xs(p^+,p^-)]$.

Similarly, the measure or rather the second derivative follows from the one in
$\CK_{\CW_A}$ under a coordinate transformation $((p')^\pm,(q')^\pm) \to (p^\pm,q^\pm)$ (defined by \er{inverselc})
\be \frac {\p^2 \tilde S(p^+,p^-; q^+,q^-)}{\p p^\pm \p q^\pm} d p^\pm \wedge d q^\pm= \frac {\p^2 S[(p')^+,(p')^-; (q')^+,(q')^-]}{\p (p')^\pm \p (q')^\pm} d (p')^\pm \wedge d (q')^\pm\,,\ee 
which has no effect on the geometry.

\begin{figure}[tbp]
\begin{center}
\includegraphics[scale=0.6]{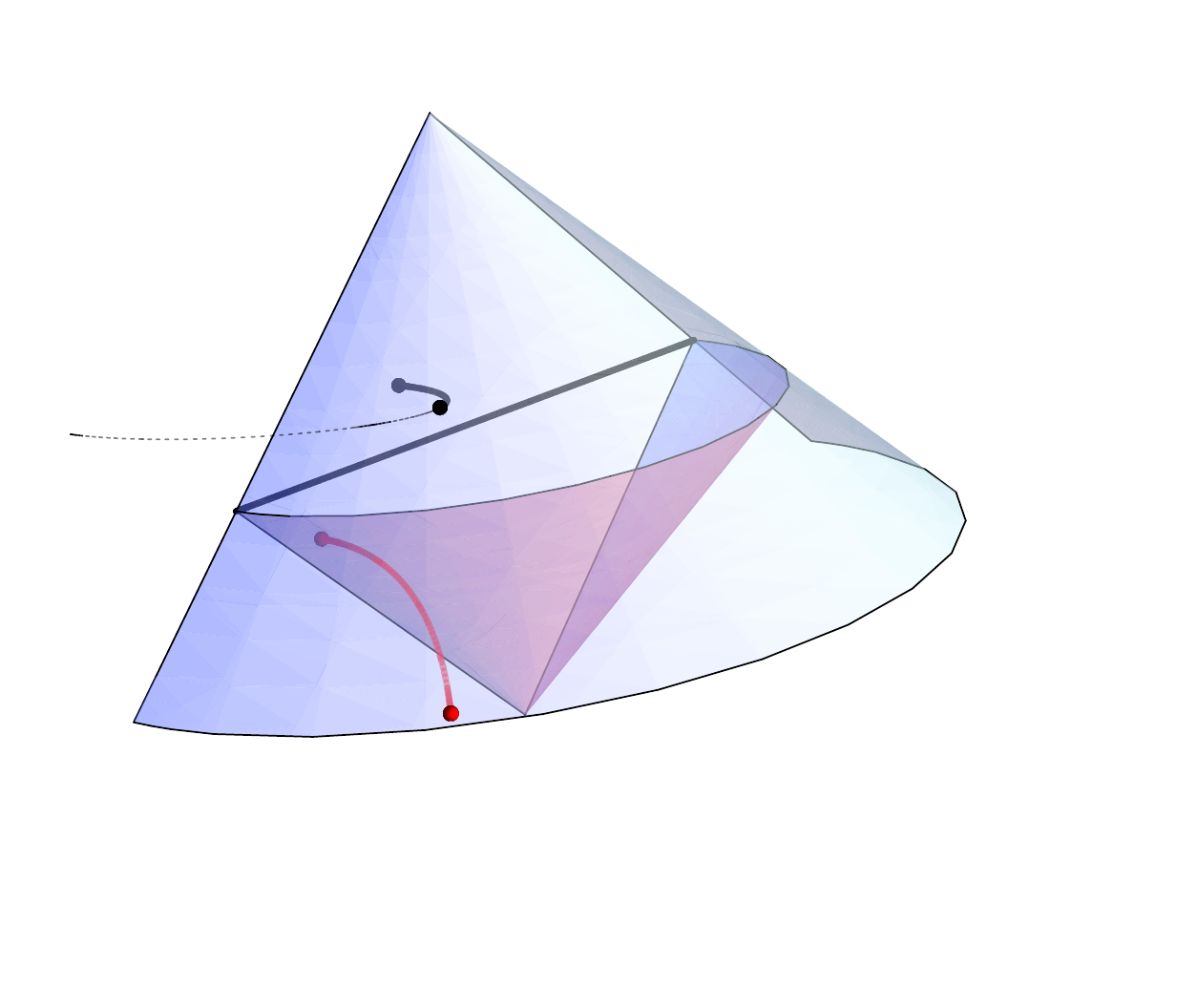}
\caption{Reflected geodesic in the entanglement wedge. The red piece $\xg_{QX'}$ is obtained by first extending the a geodesic $\xg_{PX}$ (solid black curve) and then applying the symmetry $I$ to the extended portion $\xg_{Q'X}$ (dotted black curve) outside the entanglement wedge.}
\label{causalwedge}
\end{center}
\end{figure}

Here are a few remarks before we conclude. Although the real space is LAdS$_3$,
we only consider the kinematic space of space-like geodesics, which have
natural dual in the field theory. In order for each piece of a reflected geodesic to be space-like, $P,Q$ need to satisfy
some constraints. It is also puzzling that the image $X'$ (as shown in fig.~\ref{causalwedge})
lies outside the entanglement wedge (even though it is on the extension of the light-like boundary). In our opinion, the second piece of the geodesic $\xg_{X'Q}$
shall be understood as an extension of $\xg_{PX}$. We have seen that the
kinematic space is insensitive to the details of the extension, which is
virtually the underlying reason for our construction to work. So this kind of trespassing may not be physically relevant. The bottom line is that the total length is a quantity that can be computed on the field theory side.  
 
\section{Discussions}
\label{sec:conclusion}
Motivated by the subregion-subregion duality we construct
the kinematic space of a bulk subregion using only the data that can be gleaned
from a boundary subregion $A$. For simplicity, we only consider space-like
geodesics in the Lorentzian AdS$_3$ and restrict $A$ to be a single interval. We find that a natural choice is the kinematic space of reflected geodesics.
In the special case when everything is on the $t=0$ slice of $\BH^2$, the reflected
geodesic is precisely the one minimizing the action (length) given two end
points. In a more general situation it consists of two components both of
which are spawned from a space-like geodesic $\xg_{PQ'}$ under a discrete symmetry $I$. $\xg_{PQ'}$ is a space-like geodesic crossing the boundary of the entanglement wedge $\CW_A$ at point $X$. The first component is the portion $\xg_{PX}$ of $\xg_{PQ'}$ lying inside $\CW_A$ while the other component is obtained by applying a discrete symmetry $I$ on the portion $\xg_{XQ'}$ outside $\CW_A$. In any event the length of the reflected geodesic can be obtained from the 2-point function of the state $\sqrt{\rho_A}$ on the field theory side.

In our construction, the discrete symmetry $I$ plays a central role. In three
dimensions, other solutions with negative cosmological constant looks locally
indistinguishable from pure AdS$_3$. In fact they can be obtained via
the quotient by certain discrete subgroup of conformal symmetry. In this sense, we expect our construction can be generalized
to other backgrounds (see \eg \cite{Czech:2014ppa,Zhang:2016evx,Cresswell:2017mbk} for discussions on the kinematic space of
a quotient space and also \cite{Balasubramanian:2014sra} for the relevant idea of entwinement). We would like to remind the reader that the main problem is again to find the right quantities on the field theory side to match up with the long geodesics. A good starting point would be the quotient invariant components of OPE blocks introduced in \cite{Cresswell:2018mpj}.
  
So far we have restricted ourselves to $\BH^2$ or AdS$_3$, generalization to higher dimensions is possible. Both the reflected geodesic and the correlator can be defined in general dimensions. The causal development of a ball-shaped region can be mapped to the Rindler space (with new coordinates $X^\mu$) \cite{Casini:2011kv}, where the discrete symmetry \er{inverselc} we introduced becomes $X^\pm \to - X^\pm$. The rest of the construction shall follow. It would also be interesting to study the case when $A$ consists of disjoint intervals. In fact this is the setup in which the entanglement of purification was introduced originally. It is well known that the RT surface undergoes phase transition when the
cross ratio varies. One shall be able to track the transition
via the reflected entropy. An interesting question we have in mind is to
locate the RT surface by applying the max-flow min-cut theorem (which has a lot of applications in the program of bit thread \cite{Freedman:2016zud}) on the flow
line given by the reflected geodesics. This is motivated by the picture to
interpret each geodesic as a Bell pair \cite{huang}, which fails in the multi-interval case. The issue however is resolved to some extend if the smoothness
constraint on the geodesics is lifted. 

Another interesting question is to construct the bulk operator within the entanglement
wedge. The construction in the global AdS space is based on the inverse Radon
transform. Each OPE block corresponds to a bulk operator smeared over a
geodesic, which is the so-called the Radon transform. It has been known \cite{czech1} that the inverse Radon transform gives the global HKLL formula \cite{Hamilton:2005ju,Hamilton:2006az}. However, the subregion-subregion duality implies that we can construct the bulk operator using only the reduced density matrix. We hope the new construction of the kinematic space can help to establish the inverse Radon transform within the entanglement wedge $\CW_A$ and the eventually provide to a new representation of the bulk operator associated with $\CW_A$.  

\appendix
\section{Length of a geodesic}\label{lengthgeo}
We work with the $t=0$ time slice of the AdS$_3$ space in the Poincare coordinates \er{metric}. The geodesics on this slice take the form of semi-circles.
The length of an arc on a semi-circle of radius $R$ specified by a point at the height $z=h$ and a point on the boundary ($z=0$) is given by 
\be\label{geodesich}\ell_1(h,R) = \log \left[\frac{2 R h}{\sqrt{R^2-h^2}+R}\right]\,.\ee
Practically it is more convenient to denote the arc using the boundary coordinates
$x_0$ of the $z=h$ points. We consider an interval $[-b, b]$ centered at the origin $x=0$ and a geodesic with one end point inside ($p<b$). Such a geodesic
hits the RT surface of the interval at another point $(z,x) = (\sqrt{b^2 - x_0^2},x_0)$. It can be shown that the corresponding geodesic is an arc with
radius
\[R=-\frac{b^2+p^2-2 p x_0}{2 p-2 x_0}\,.\]
Plugging this back into \er{geodesich}, we obtain the length of this arc
as
\be\label{geodesicl1}\ell(p,x_0) = \log \left[\frac{b^2+p (p-2 x_0)}{\sqrt{b^2-x_0^2}}\right]\,.\ee

Let us now go beyond the $t=0$ slice and put an end point $P$ at $(z,t,x) = (0, p_2, p_1)$ and the other $X$ at $(\sqrt{(b-y_0)^2-x_0^2},y_0, x_0)$. We note that the second end point lies on the light cone in the bulk emitted
from the tip $(0, b, 0)$ of the causal diamond. Again the geodesic is an
arc whose radius can be determined by the following equation
\[(b-y_0)^2+\left(R-\sqrt{(p_1-x_0)^2-(p_2-y_0)^2}\right)^2-x_0^2=R^2\,,\]
which gives
\[R= \frac{b^2-2 b y_0+p_1^2-2 p_1 x_0-p_2 (p_2-2 y_0)}{2 \sqrt{(p_1-x_0)^2-(p_2-y_0)^2}}\,.\]
The variables in equation \er{geodesich} are boost invariant and hence it
remains valid even when the two end points have different time coordinates.
The length of the geodesic $\xg_{PX}$ is then given by
\be
\label{geodesicl2}
\ell(p_1, p_2; x_0, y_0)=\log \left(\frac{b^2-2 b y_0+p_1^2-2 p_1 x_0-p_2^2+2 p_2 y_0}{\sqrt{b^2-2 b y_0-x_0^2+y_0^2}}\right)\,.\ee

\section{$\br \CO(p) \xD_A^{1/2}  \CO(q) \ke$ and reflected geodesic}
\label{connectionmodular}
It was proposed in \cite{Faulkner:2018faa} that the correlator $\br \CO(p) \xD_A^{i s}  \CO(q) \ke$ corresponds to two geodesics which are connected at a certain point on the RT-surface $\xg_A$ and are related by a boost that leaves $\xg_A$ invariant. We will show that for the special case of $s= -\frac i 2$ and $x,y$ both on $t=0$ slice,  the configuration becomes a reflected geodesic. First of all, $\br \CO(p) \xD_A^{i s}  \CO(q) \ke$ is the correlator with an insertion of the modular operator. For a certain subregion $A$ (and a certain state $\Psi$, which in our case is just the vacuum), we have this
operator $S_A$ that takes any operator in $A$ to its Hermitian conjugate.
This operator can be polar decomposed into an anti-unitary part $J_A$  and
a Hermitian part $\xD^{1/2}$ which is the modular operator (see \eg \cite{Witten:2018lha} for a review)
\[S_A a\st{\Psi} = a^\dagger \st{\Psi},\quad S_A = J_A \xD_A{}^{1/2}\,. \]

Following \cite{Faulkner:2018faa}, the correlator $\br \CO(p) \xD_A^{is}  \CO(q) \ke$ can be computed holographically using two geodesics that meet somewhere
(denoted by a bulk point $X$) on the RT surface $\xg_A$ 
\[
\br \CO(p) \xD^{is}  \CO(q) \ke \simeq \exp \lb - m [\ell(p,X)+\ell(X,q)]\rb
\]
where $m$ is the mass of the bulk field dual to $\CO$. Their tangent vectors at the transition point are related by a boost at the bulk point $X$ ($\pm$ to denote the light-cone components)
\[n_i' = n_i,\quad (n_+',n_-') = (e^{-2\pi s} n_+, e^{+2\pi s} n_-)\]

We consider the case with all the geodesics lying entirely within the constant time slice of $A$ and $\xg_A$. It can be checked that
for the case of $\xD_A^{1/2}$ the matching boundary condition is precisely
what makes the total length minimized. To be precise, the condition is that the
 component orthogonal to $\xg_A$ is reversed while the tangent components
 are kept invariant, which is essentially the condition for a reflection. As
we learn in optics, such a configuration is reached via action principle ($\xd \ell(p,X)+\xd \ell(X,q) = 0$)\footnote{The variation is with respect
to the coordinates on $\xg_A$. The derivatives of the action (length) with respect to the position coordinates give momentum and hence the action principle
implies the continuity of the longitudinal (tangent to $\xg_A$) momentum.
There is only one normal direction in our setup and hence it is trivial that
the normal component should be reversed.}. In summary the length of a reflected
geodesic is given by the correlator with modular operator $\xD_A^{1/2}$.

\section*{Acknowledgments}
This work is supported by the NSFC Grant No.11947301, the NWU Grant No.0115/338050048, No.0202/334041900012 and the Double First-class University Construction Project of Northwest University.

\bibliography{bio}

\end{document}